\documentclass[final,1p,times]{elsarticle}

\usepackage{graphicx}

\usepackage{amssymb}

\journal{Nuclear Physics A}

\begin{document}

\begin{frontmatter}



\title{The $(K^-,p)$ reaction on nuclei with in-flight kaons}


\author{A. Ramos$^a$, V. K. Magas$^a$, J. Yamagata-Sekihara$^{b,c}$, 
S. Hirenzaki$^d$ and E. Oset$^c$}

\address[ECM-UB]{Departament d'Estructura i Constituents de la Mat\`eria 
and Institut de Ci\`encies del Cosmos, \\
Universitat de
Barcelona, 08028 Barcelona, Spain}
\address[YITP]{Yukawa Institute for Theoretical Physics, Kyoto University, Kyoto 606-8502, Japan}
\address[FT-UV]{Departamento de F\'{\i}sica Te\'orica and IFIC, Centro Mixto Universidad de Valencia-CSIC, \\
Institutos de Investigaci\'on de Paterna, Apartado 22085, 46071 Valencia, Spain}
\address[NWU]{Department of Physics, Nara Women's University, Nara 630-8506, Japan}

\begin{abstract}

We perform a theoretical study of the spectrum of protons with kinetic energies
of around 600 MeV, emitted following the interaction of 1 GeV/c kaons with
nuclei. A recent experimental analysis of this $(K^-,p)$ reaction on $^{12}$C,
based on the dominant quasielastic process, has suggested a deeply attractive
kaon nucleus potential. Our Monte Carlo simulation considers, in addition,
the one-and two-nucleon $K^-$ absorption processes producing hyperons that
decay into $\pi N$ pairs. We find that this kaon in-flight reaction is not well suited to
determine the kaon optical potential due, essentially, to the limited
sensitivity of the cross section to its strength, but also to unavoidable
uncertainties from the coincidence requirement applied in the experiment.
A shallow kaon nucleus optical potential obtained in
chiral models is perfectly compatible with the observed spectrum.

\end{abstract}


\end{frontmatter}


\section{Introduction}
\label{intro}

The study of the interactions of antikaons
with nuclei has received much
attention in past years. Although kaonic-atom data favor an attractive
 $K^-$-nucleus interaction \cite{friedman-gal,gal}, the discussion
centers on whether it can accommodate deeply
bound antikaon states which could be observed in direct reactions.
Potentials based on underlying chiral dynamics of the kaon-nucleon interaction
\cite{lutz,angelsself,schaffner,galself,Tolos:2006ny} show a
moderate attraction of the order of 60 MeV at nuclear matter density and 
have a sizable imaginary part, which would rule out
the experimental observation of peaks. Nevertheless,
the theoretical shallow potentials
reproduce satisfactorily data of kaonic atoms \cite{okumura}, and
best fits including an additional
 phenomenological component \cite{baca} indicate 
 deviations of at most
 20\% from the theoretical potential of \cite{angelsself}. Recently, 
 the lightest ${\bar
 K}NN$ system has also been the subject of 
 strong debate \cite{akaishi:2007cs,shevchenko,hyodo,sato}.
 
At the other extreme, highly attractive phenomenological potentials having
a strength of about 600 MeV at the center of the nucleus, leading
to nuclear densities ten times that of normal nuclear matter,
and favoring the existence of deeply bound kaon nuclear
states, have been advocated
 \cite{akaishi:2002bg,akainew}, but have
 received strong criticisms \cite{toki,Hyodo:2007jq,npangels}.
Furthermore, the claimed experimental evidences of deeply bound 
states using {\it stopped} kaon reactions at KEK \cite{Suzuki:2004ep}
and FINUDA \cite{Agnello:2005qj,:2008chb} have to be looked at with caution, 
especially
after the reanalysis of the KEK  experiment \cite{Sato:2007sb} which makes it
consistent with the spectra measured in \cite{:2007ph}, interpreted
there on the basis of the two-nucleon absorption mechanism pointed out in
\cite{toki}. The calculations of Refs.~\cite{Magas:2006fn,Magas:2008bp} explain
also the FINUDA peaks \cite{Agnello:2005qj,:2008chb} in terms of conventional
two- and three-nucleon absorption mechanisms, consistently with  recent KEK
data \cite{Suzuki:2007pd,Suzuki:2007kn}.

There is yet another experiment, measuring fast protons
emitted from the reaction of {\it in flight} kaons with
nuclei \cite{Kishimoto:2007zz},
from which evidence for a very strong kaon-nucleus potential, with a depth
of the order of 200 MeV, was claimed. The data was analyzed in terms
of the Green's function method 
of Refs.~\cite{zaki,Yamagata:2007cp,YamagataSekihara:2008ji}
considering only the dominant quasielastic
$K^- p \to
K^- p$ process. In this contribution we will 
show that one- and two-nucleon
absorption reactions also contribute to the spectrum,
the shape of which is, moreover,  
strongly affected by the experimental coincidence requirements, 
making this reaction not well suited to extract information on 
the depth of the kaon optical potential.

\section{Monte Carlo simulation of the \mbox{{\boldmath $(K^-,p)$}} reaction}
\label{monte-carlo}

Our Monte Carlo simulation of the $(K^-,p)$ reaction \cite{magasnew,ISMD2009} 
considers the nucleus  as
a density distribution of nuclear matter, through which the kaon and all  other
produced particles propagate. As sources of fast protons we consider the
quasielastic processes ($\bar{K} N \to \bar{K} N$), as well as kaon absorption
by one nucleon ($\bar{K}  N \to \pi \Lambda$, $\bar{K}  N \to \pi \Sigma$) or two
nucleons ($\bar{K} N N\to  \Sigma N$ and $\bar{K} N N\to  \Lambda N$), 
followed by the weak decay of the $\Sigma$ or the $\Lambda$ into $\pi N$ pairs.
Each reaction occurs
according to the probability $\sigma_i \rho(r) \delta l$ ($i={\rm qe, 1N,
2N}$), where $\rho(r)$ is the local density and $\delta l$ a small enough step
size.  The values of the cross sections, taken from the Particle Data Group
\cite{PDG}, are explicitly listed in Ref.~\cite{magasnew}.

In case of a quasielastic collision, the direction of the scattered kaon and
nucleon is determined according to the experimental differential cross section,
checking always that the size of the nucleon momentum is larger than
that of the local Fermi sea. The nucleon then propagates through the
nucleus colliding with other
nucleons, losing energy, changing its direction, and generating new secondary
nucleons. The rescattered kaon is also followed, letting it experience different
interaction mechanisms (scattering or absorption) according to their respective
probabilities.  
In one-nucleon absorption processes of the type $K^- N \to \pi \Lambda$ and 
$K^- N \to \pi \Sigma$, we let the $\Lambda$ or the $\Sigma$ propagate
undergoing quasielastic collisions with the nucleons and, once
they leave the nucleus, they are allowed to decay weakly into $\pi N$ pairs,
providing in this way a source of protons in the energy region of interest.
The simulation also accounts for two-nucleon absorption processes $K^- NN \to
\Lambda N$ or $K^- NN \to \Sigma N$ with all possible charge combinations. 
According to $^4$He data \cite{Katz:1970ng}, the
probability per unit length for two-nucleon absorption, 
$\mu_{K^-NN}\equiv C_{\rm abs}\rho^2$, is taken
to be 20\% that of one-body absorption, which determines a value 
$C_{\rm abs} \simeq 6$ fm$^5$. The two-body absorption reactions
provide a double source of fast protons, those produced directly in the
absorption process and those coming from the decay of the
hyperon into
$\pi N$ pairs.

We also implement the effect of a kaon optical potential, $V_{\rm opt}={\rm Re}\,
V_{\rm opt} + {\rm i}~ {\rm Im}\, V_{\rm opt} $, which will influence the kaon
propagation, especially after a high momentum transfer
quasi-elastic collision when the kaon will acquire a relatively low momentum.

At the end of the simulation, we  keep the events containing
a proton in a kinetic energy range of  $500-700$ MeV,
within an angle of 4.1 degrees in the lab frame, 
as in the experiment. 
To facilitate
comparison with experiment, the missing invariant mass 
is translated into a kaon binding energy, $E_B$,
according to:
\begin{equation}
\sqrt{(E_K+M_{^{12}{\rm C}}-E_p)^2-(\vec{P}_p - \vec{P}_K)^2} = M_{^{11}{\rm B}} + M_{K}
-E_B \,,
\label{B_E}
\end{equation}
where $E_p,\vec{P}_p$ are the energy and momentum of the observed proton and
$E_K,\vec{P}_K$ are the energy and momentum of the initial kaon.

\section{Results and Discussion}
\label{results}

Our results for an optical potential
$V_{\rm opt}=(-60,-60)\rho/\rho_0$ MeV are shown in Fig.~1,
including only quasi-elastic
reactions (dash-dotted line) and considering also one- and two-nucleon
absorption processes (solid line). 
A non-negligible amount of
strength is gained in the region of ``bound kaons" due to the new mechanisms.
Although not shown separately, we find that
one-nucleon absorption and multi-scattering processes
contribute to the region $-E_B > -50$ MeV whereas the 
two-nucleon absorption reactions contribute
to all values of $-E_B$, starting from values as low as 
$-300$ MeV.\\

\begin{minipage}[t]{.45\linewidth}
\includegraphics[width=\textwidth]{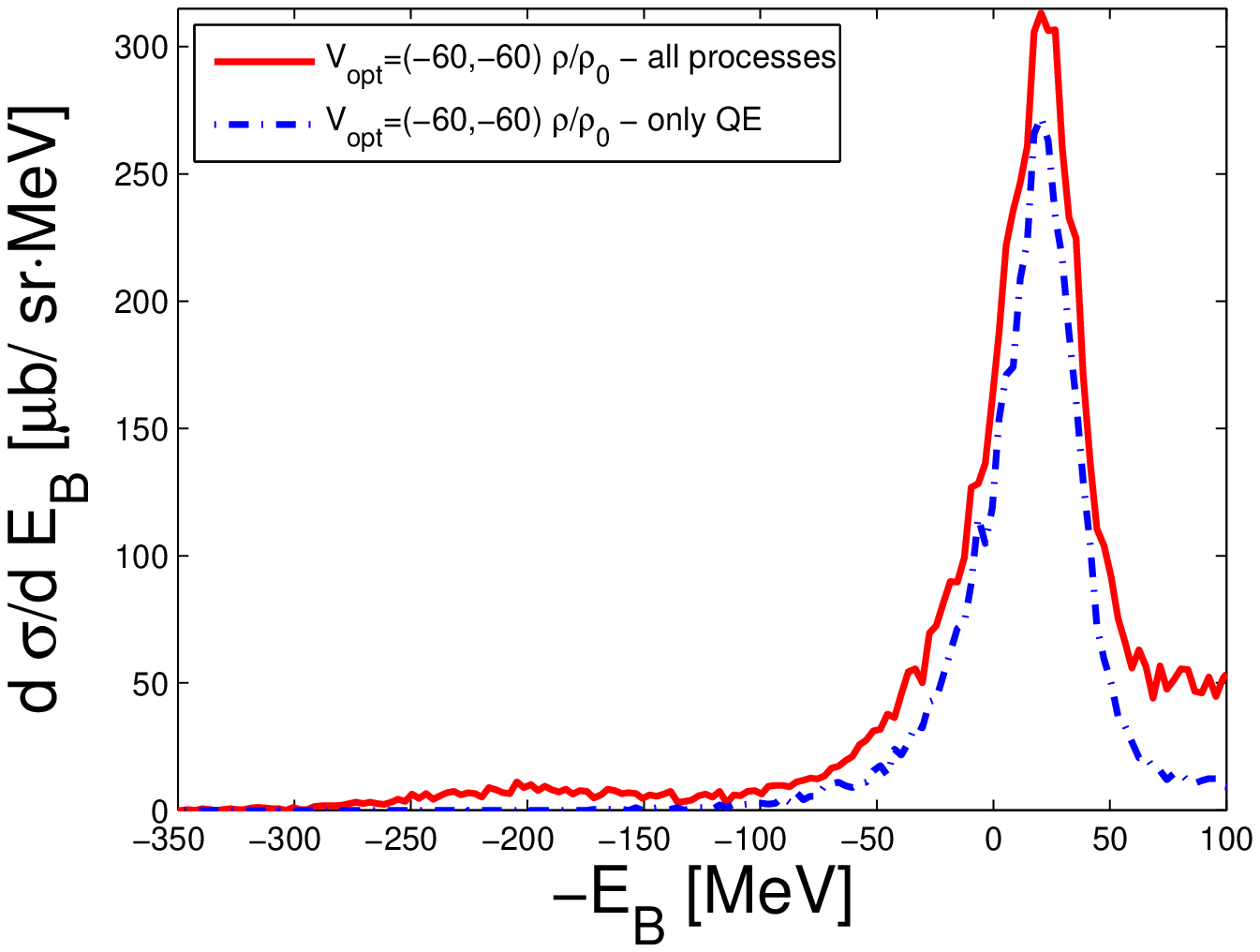}\\
Fig.1. Proton spectra for quasi-elastic processes (dash-dotted line),
  and including all processes (solid line), using
 $V_{\rm opt}=(-60,-60)\rho/\rho_0$ MeV.
\end{minipage}
\hfill
\begin{minipage}[t]{.45\textwidth}
\includegraphics[width=\textwidth]{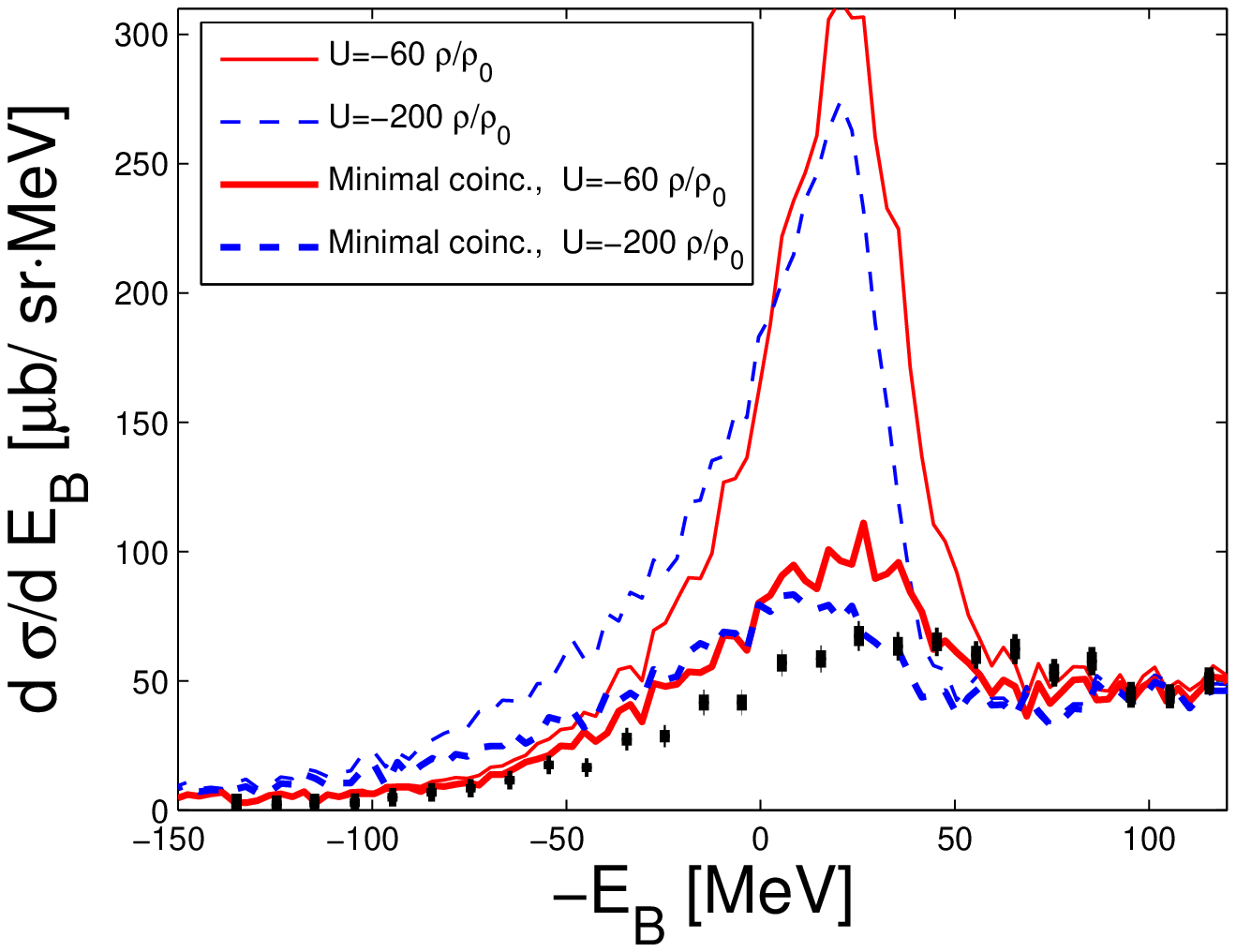}\\
Fig. 2. Proton spectra without (thin lines) and with (thick lines)
the minimal
coincidence requirement, using
$V_{\rm opt}=(-60,-60)\rho/\rho_0$ MeV (solid lines) and  
$V_{\rm opt}=(-200,-60)\rho/\rho_0$ MeV
 (dashed lines). Experimental data are 
  from \cite{Kishimoto:2007zz}.
\end{minipage}\\

We must keep in mind that the outgoing forward protons were measured in
coincidence with at least one charged particle in the decay counters
surrounding the target \cite{Kishimoto:2007zz} and 
the analysis
assumed the shape of the spectrum not to change under that requirement.
Whereas a detailed simulation of these experimental
conditions is prohibitive, we can at least study their 
consequences by applying
a minimal coincidence requirement \cite{magasnew,ISMD2009},
which  eliminates the events that, for sure, will not produce a
coincidence. These are the ones that, after a primary quasi-elastic collision
producing a fast forward proton and a backward kaon, neither particle suffer
any further reaction and, therefore, no charged particle will have the chance 
of hitting the decay counters. This minimal coincidence 
requirement changes the shape of the spectrum considerably, as seen in
Fig.~2 upon comparing
the bare spectrum obtained with
a kaon potential depth of 60 MeV (solid thin line) with that obtained after the
coincidence cut (solid thick line). The distribution becomes wider
because many ``good events" generated from the dominant quasielastic processes
are eliminated. The figure also shows the spectra
corresponding to a potential depth of 200 MeV, before (dashed thin line) and
after the minimal coincidence cut (dashed thick line). We clearly see that 
the sensitivity of the bound region to the optical potential is
essentially lost when the coincidence requirement is applied. 

We finally
note that our implementation of the experimental conditions
should actually lead to a spectrum that overshoots
the data, as it keeps some events that might not produce a coincidence signal.
The amount of discrepancy should be smaller or inexistent in the continuum 
region, populated by lower momentum protons produced in
many particle final states, which have a better chance of hitting the decay
counters. Having this in mind, the spectrum obtained with a kaon nucleus
potential of $V_{\rm opt}=(-60,-60)\rho/\rho_0$ MeV is compatible with the
experimental data, as one can see in Fig.~2.

Our results
demonstrate the limited capability of the $(K^-,p)$ reaction with in-flight
kaons to infer the depth of the kaon optical potential. 
On the one hand, there are more processes beyond quasielastic reactions that
populate the spectrum in the region of interest and, on the other hand,
large uncertainties are introduced when trying to simulate the conditions 
of the experimental set up \cite{Kishimoto:2007zz}. 
Contrary to what it is assumed in the analysis of Ref.~\cite{Kishimoto:2007zz},
we have clearly seen
that the spectrum shape is affected by the required coincidence.
Certainly, the bare spectrum would be a much 
more valuable observable to learn about the kaon nucleus optical potential.



\end{document}